\documentclass[aps,pra,showpacs,amsmath,twocolumn,superscriptaddress]{revtex4-1}
\usepackage{amsmath,amssymb,graphicx}
\usepackage[final]{hyperref}
\usepackage{booktabs}

\hypersetup{
	colorlinks=true,       
	linkcolor=blue,          
	citecolor=blue,        
	filecolor=magenta,      
	urlcolor=blue
	}

\begin{document}

\title{Spatial Multi-Mode Structure of Atom-Generated Squeezed Light}

\author{Mi Zhang}
\affiliation{Department of Physics, College of William $\&$ Mary, Williamsburg, Virginia 23187, USA}
\author{R. Nicholas Lanning}
\author{Zhihao Xiao}
\author{Jonathan P. Dowling}
\affiliation{Hearne Institute for Theoretical Physics and Department of Physics $\&$ Astronomy, Louisiana State University, Baton Rouge, Louisiana 70803, USA}
\author{Irina Novikova}
\author{Eugeniy E. Mikhailov}
\email[eemikh@wm.edu]{}
\affiliation{Department of Physics, College of William $\&$ Mary, Williamsburg, Virginia 23187, USA}

\date{\today}

\begin{abstract}
We investigated the spatial distribution of quantum fluctuations in a squeezed vacuum field, generated via polarization self-rotation (PSR) interaction of an ensemble of Rb atoms and a strong near-resonant linearly polarized laser field. We found that the noise suppression is greatly effected by the transverse profile of a spatial mask, placed in both the squeezed field and the local oscillator, as well as its position along the focused beam near the focal point. These observations indicate the spatial multi-mode structure of the squeezed vacuum field. We have developed a theoretical model that describes the generation of higher-order Laguerre-Gauss modes as a result of PSR light-atom interaction. The prediction of this model are in a good qualitative agreement with the experimental measurements.
\end{abstract}
		

\maketitle

\section{Introduction}
The ability to control the spatial mode composition of optical fields promises practical applications for many quantum technologies~\cite{polzik_book}. 
For example, multi-mode squeezing and entanglement can be used for quantum imaging~\cite{lettSci08,Dowling2008}, quantum information  multiplexing~\cite{PhysRevLett.98.083602,Torres2007NaturePhy}, hyper-spectral encoding, etc.
At the same time, the presence of several spatial modes may be detrimental
for quantum-enhanced measurements~\cite{fabre:sfo-00270537}. If the
quadrature angle, corresponding to the maximum quantum noise suppression,
is different for different spatial modes, the overall detected noise may be
increased, even if each individual spatial mode is squeezed. However, if the proper mode structure is identified, it may be possible to extract  a single  squeezed mode
and eliminate  others by means of an efficient spatial mode sorter~\cite{Berkhout2010PRLoamsorting} or to tailor the apparatus~\cite{FabremultimodePRA2012} to further boost the signal-to-noise ratio
(SNR). In this way, we would be able to produce single, maximally squeezed, higher-order modes upon demand.

In this paper, we investigate the spatial properties of the resonant optical field after interaction with dense Rb vapor under the conditions for polarization self-rotation (PSR)~\cite{Davis:92,budkerPRA01,novikova02JMO}. It is known that in the case of a linearly polarized input field, the noise in the orthogonal polarization is modified, leading to the generation of squeezed vacuum~\cite{matsko_vacuum_2002, ries_experimental_2003, mikhailov2008ol}. Several experiments demonstrated up to 3~dB of quantum noise suppression with this scheme~\cite{lezama2011pra,grangier2010oe,mikhailov2012sq_magnetometer}. It is still an open question  why better quantum noise suppression cannot be experimentally achieved, in contrast to the predictions of the theory~\cite{matsko_vacuum_2002,lezama_numerical_2008,mikhailov2011jmo}. Here we present a study of the spatial mode structure of the output optical fields in the PSR squeezing process, that gives strong evidence of its multi-mode nature. In particular, we carried out a simple sorting of the spatial modes, elucidating the mode structure of the pump and the squeezed field. Furthermore, we developed an intuitive theoretical description of the  higher-order spatial modes  that qualitatively agrees with the experiment.


\begin{figure}[b]
	\includegraphics[width=1.0\columnwidth]{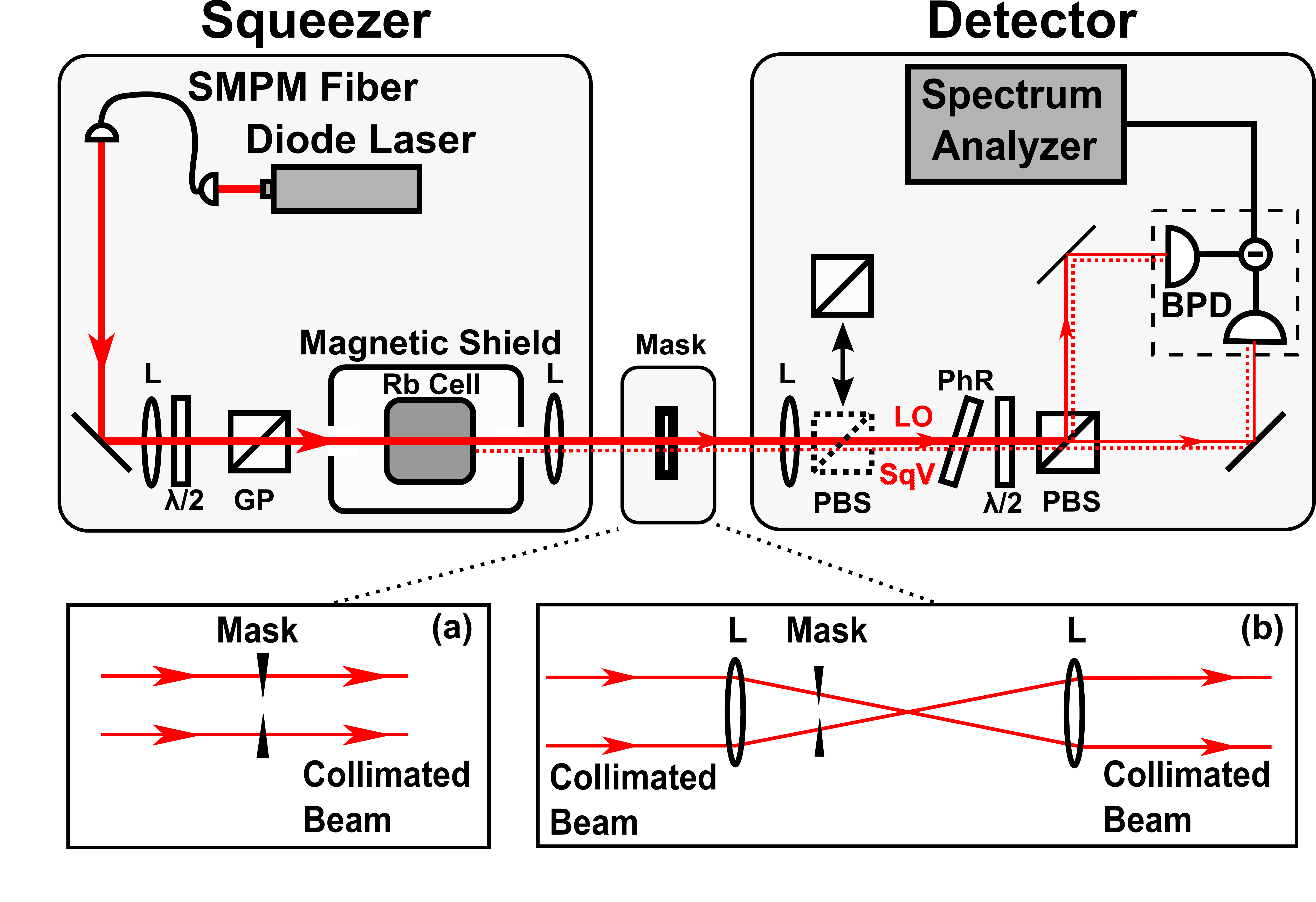}
	\caption{
		\label{fig:setup}(Color online)
		Experimental setup.
		SMPM is a single-mode polarization-maintaining fiber,
		$\lambda/2$ is half-wave plate,	
		GP is Glan-laser polarizer,
		PBS is a polarizing beam splitter,
		PhR is a phase-retarding wave plate,
		and BPD is a balanced photodetector.
		In the experiment a mask was inserted either in the collimated
		beam (a) or in the focused beam (b).
	}
\end{figure}

\section{Experimental apparatus}
The experimental setup for squeezing generation, depicted at the Fig.~\ref{fig:setup}, is similar to the one used in our previous experiments~\cite{mikhailov2013ol_vortex}. An extended cavity diode laser was tuned by approximately 100~MHz to the red of the $5^{2}S_{1/2}, F=2 \rightarrow 5^{2}P_{1/2},F^{\prime} = 2$ transition of ${}^{87}$Rb ($\lambda \simeq$ 795~nm). The laser output was spatially filtered by passing it through a single-mode-polarization-maintaining (SMPM) fiber, linearly polarized using a Glan-laser polarizer (GP) and focused into a 7.5~cm long cylindrical Pyrex cell with isotropically enriched $^{87}$Rb vapor and no buffer gas. The waist of the focused beam (diameter $100~\mu$m at $1/e^2$ intensity level) was located 6.5~cm from the front of the cell. The cell was mounted inside three layer $\mu$-metal magnetic shielding, and maintained at a constant temperature of $66^{\circ}$C, corresponding to a Rb density of $5.4\times10^{11}$~cm$^{-3}$.

The interaction of a strong laser field with the $^{87}$Rb vapor modifies the vacuum fluctuations in the orthogonal polarization. In order to measure quadrature noise, we rotated the polarization of the two fields by 45$^\circ$ and mixed them on a polarization beam splitter (PBS) before sending them to a balanced photodiode (BPD). We thus realized a homodyne detection in which the strong laser field served as the local oscillator (LO)~\cite{lezama2011pra, mikhailov2012sq_magnetometer}. The relative phase (quadrature angle) of the two fields was adjusted by tilting the phase-retarding (PhR) plate (a birefringent quarter-wave plate with a crystal axis set parallel to the LO polarization). All the measurements reported below were performed at 1~MHz detection frequency with $100$~kHz resolution and 30~Hz video bandwidths. To calibrate the SQL (Standard Quantum Limit, i.e., shot-noise) level, we inserted an additional PBS into the beam after the cell, which reflects the squeezed field and passes only the LO field. We detected noise suppression of $1.9\pm0.2$~dB below the SQL level in the maximally squeezed quadrature and 11~dB of anti-squeezing in the orthogonal quadrature (see Fig.~\ref{fig:iris_and_dot_and_combined} at 100\% transmission level), which is quite typical for these kinds of atomic squeezers.

\section{Spatial Properties of the Collimated Squeezed Beam}

All previous experimental and theoretical analysis of PSR squeezing assumed an identical single spatial mode for both the strong pump and vacuum fields, with either fundamental Gaussian~\cite{lezama2011pra, mikhailov2012sq_magnetometer} or
Laguerre-Gaussian~\cite{mikhailov2013ol_vortex} transverse profiles. However, as we show below, the spatial composition of the squeezed field mode is more complex.

In order  to investigate  the spatial  property of  the squeezed  field, we inserted different radially symmetric spatial masks into the collimated beam after the Rb cell (shown in Fig.\ref{fig:setup}(a)). We classify the beam masks as irises or disks, depending on whether they block the outer or inner part of the laser beam. We used an adjustable iris, and a set of fixed sizes disks. These masks were inserted into the collimated part of the laser beam after interaction with the atoms, making sure that the masks are properly centered. Since the masks reduced the LO power, we carefully recalibrated the shot-noise level for every mask and adjusted the phase retarding plate to track the maximally squeezed and anti-squeezed quadrature noises.

\begin{figure}[h]
   \includegraphics[width=1.0\columnwidth]{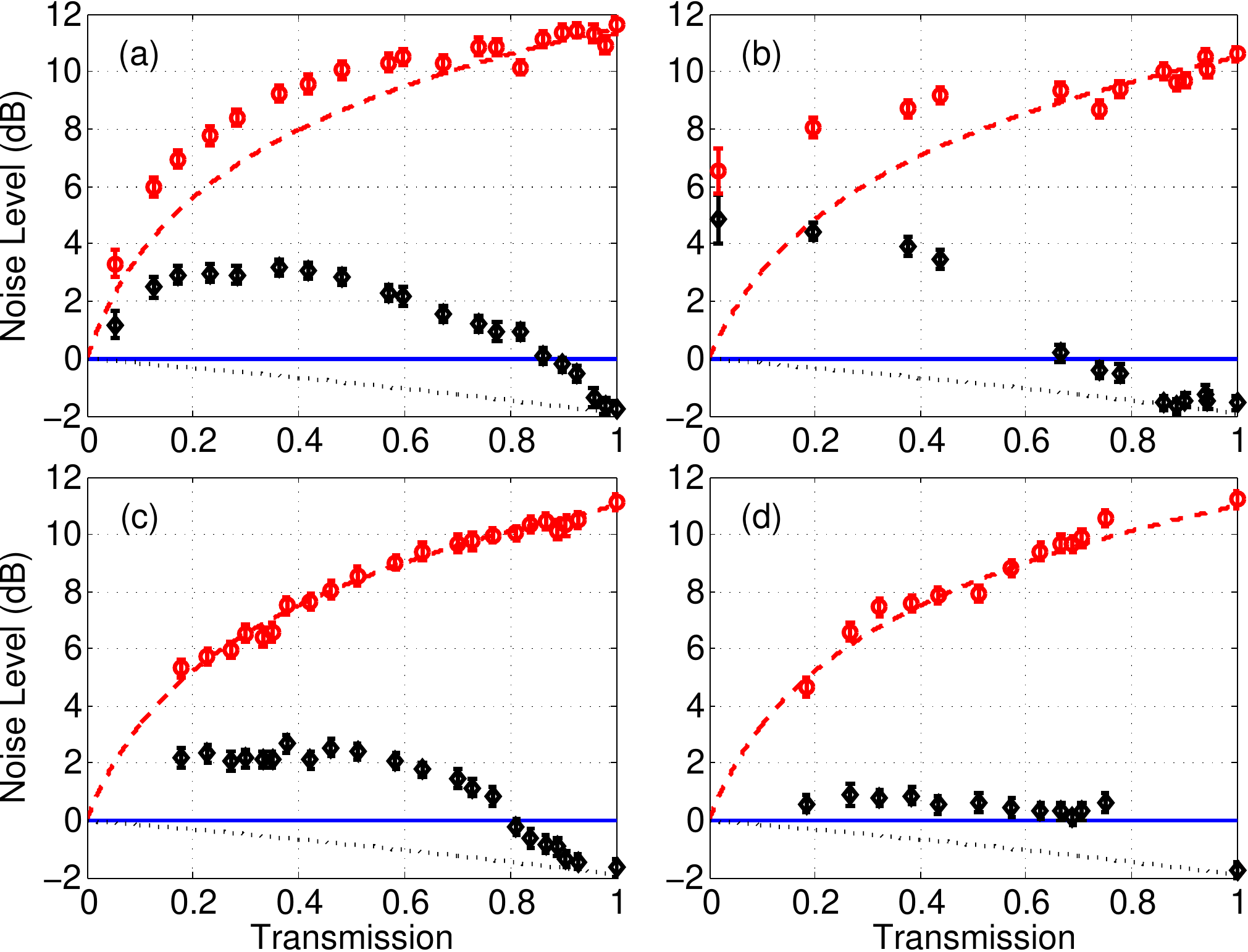}
	\caption{
		\label{fig:iris_and_dot_and_combined}
		(Color online)
		Measured minimum (diamonds) and maximum (circles) quadrature noise when the laser beam is partially blocked by (a) an iris mask, (b) a disk mask, (c,d) the combined 				masks, formed by both iris and the disk. For all measurements the horizontal axis indicates the fraction of LO intensity, transmitted through the mask.
		In (c) the central disk alone blocked 8\% of the LO	power and in (d) the power loss was 25\%.~ Dashed and dotted lines indicate the prediction of the single-mode model, described by Eq.~(\ref{eq:beamsplitter}).
		The zero of the vertical axis corresponds to the shot-noise	noise level .
	}
\end{figure}

The modifications of the quantum noise by different types of masks are presented in Fig.~\ref{fig:iris_and_dot_and_combined}. Based on a single-mode description of the optical field, one would expect the changes in the detected quantum noise in all cases to depend only on the total optical transmission $T$, and to be accurately described by the beam-splitter expression:
\begin{equation}
	\label{eq:beamsplitter}
	SqV_{\rm{out}}=10 \cdot \rm{log}_{10}[T\cdot10^{SqV_{in}/10}+(1-T)],
\end{equation}
where $\ SqV_{\rm{in,out}} $ are the quadrature noise measured in dB before and after the mask. Clearly, the experimentally measured noise values deviate significantly from the naive expected dependencies; shown in Fig.~\ref{fig:iris_and_dot_and_combined} as dashed and dotted lines, indicating non-trivial spatial correlations~\cite{MarinoPhysRevLett08,MarinoEJPD2012}. For example, even small losses of about 10\% (for the iris mask) and 30\% (for the disk mask) bring the squeezed quadrature noise significantly above shot noise. Moreover, for the disk mask, even at small transmission ($T <5$\%), the quantum noise in both quadratures is more than 5~dB above shot noise, see Fig.~\ref{fig:iris_and_dot_and_combined}(b). In contrast, Eq.~(\ref{eq:beamsplitter}) predicts that in all cases we should expect the noise to approach the SQL monotonically from below, never exceeding the shot noise level.

To gain additional insight about the spatial distribution of the squeezed vacuum field, we looked at the noise of a ring-like slice of the laser beam. To do this we constructed a mask consisting of a fixed size opaque disk and a variable size iris.  Fig.~\ref{fig:iris_and_dot_and_combined}(c,d) represents modifications of transmitted quantum noise by such masks where the fixed disk sizes are characterized by 8\% and 25\% blockage of LO power. Again, Fig.~\ref{fig:iris_and_dot_and_combined}(c,d) shows that we were not able to improve the measured noise suppression below that of the unobstructed beam, even though the anti-squeezing noise followed the uniform loss model much better. Perhaps, such combined masks were able to block especially noisy spatial modes.

The above observations suggest that the generated squeezed field consists of several spatial modes, some of which are ``noisier'' than others.  These modes are expected to be radially symmetric due to the cylindrical symmetry of
our setup. The higher-order Laguerre-Gaussian (LG) modes~\cite{Siegman_book} are thus  natural candidates to use for  the decomposition of the LO and the squeezed beam. Such models are written as:
\begin{equation} \label{eq:LGmodes}
 \begin{split}
  	u_{l,p}(\vec{r})= \dfrac{C_{l,p}}{w(z)} e^{- \frac{r^{2}}{w(z)^{2}}}e^ { -\frac{ikr^{2}z}{2(z^{2}+z_{R}^{2})} }  \big( \dfrac{\sqrt{2}r}{w(z)} \big)^{|l|} \\
	\times  \: L_{p}^{|l|}
	\big( \dfrac{2r^{2}}{w(z)^{2}} \big)  e^{il\phi} e^{i(2p+|l|+1)\arctan(z/z_{R})} \end{split},
\end{equation}
where $l$ is the azimuthal index and $p$ is the radial index for each mode, $ C_{l,p}=\sqrt{2p! / \pi(|l|+p)!}$ is a normalization constant, $w_{0}$ is the beam waist, $ w(z)=w_{0}\sqrt{1+(z/z_{R})^2}$ is the width of the optical field, $L_{p}^{|l|}$ are the generalized Laguerre polynomials, $z_{R}=\pi \omega_{0}^{2}/\lambda$ is the Rayleigh range, and $k=2\pi/\lambda$ is the wave
number. Because of the conservation of the optical angular momentum, we expect that only the modes with $l=0$ can be generated via the PSR process.

\section{Semi-Classical Theory of the Spatial Multi-Mode Generation}
Since the previous theory~\cite{matsko_vacuum_2002,lezama_numerical_2008} was performed in the plane-wave approximation, we develop a model that accounts for the possibility of the higher-order spatial modes. Treating the light classically, we start with the inhomogeneous wave equation
 \begin{equation}\label{eq:inhomogeneous wave equation}
 \nabla^{2}\textbf{E}-\frac{1}{c^{2}}\frac{\partial^{2}\textbf{E}}{\partial t^{2}}=	\frac{1}{\epsilon_{0} c^{2}} \frac{\partial^{2}\textbf{P}}{\partial t^{2}},
 \end{equation}
where $ \textbf{P}=\frac{N}{V}\langle\hat{\textbf{d}}\rangle $ is the polarization induced in the medium, $ \langle \hat{\textbf{d}} \rangle = \sum_{m,n} \mu_{mn} \rho_{mn}$ is the expectation of the dipole operator in an atomic level basis $\{|n\rangle\}$, and $\mu_{mn}$ are dipole moment matrix elements. We use the density matrix expression for the polarization, along with the slowly varying envelope and paraxial wave approximations, to transform Eq.~(\ref{eq:inhomogeneous wave equation}) into a propagation equation
in terms of the envelope functions $\tilde{E} $ and $ \tilde{\rho}$:
  \begin{equation}
(\frac{\partial}{\partial z}-\frac{i}{2k}\nabla^{2}_{\perp})\tilde{E}=\frac{i k}{2 \epsilon_{0}} \frac{N}{V}\sum_{m,n} \mu_{mn}\tilde{\rho}_{mn}.
 \end{equation}
Next we apply a simple model in which we treat the D1 line of $^{87}$Rb  as a double-$\Lambda$ scheme\cite{Lett2012PRLDoubleL}. We treat the input field as a superposition of two circularly polarized fields characterized by Rabi frequencies $\Omega_+$ and $\Omega_-$, corresponding to right- and left-circular polarizations. We solve for the density matrix elements, and since our pump field is in fact linearly polarized, we convert to the linear polarization basis and distinguish the propagation equations
for the two envelope functions in terms of the Rabi frequencies:
 \begin{equation}\label{eq:Rabi frequency propagation}
 	\hat{L}\:\Omega_{x,y}=-2\kappa \Omega_{x,y} \frac{|\Omega_{x,y}|^{2}}{|\Omega|^{4}}(\gamma_{0}+2i\frac{|\Omega_{x,y}|^{2}}{\Delta})	
 \end{equation}
where $\hat{L} \equiv (\frac{\partial}{\partial z}-\frac{i}{2k}\nabla^{2}_{\perp})$, $k$ is the wave number, $\kappa$ is
the coupling constant, $\gamma_{0}$ is the decay rate, $|\Omega|^{4} \equiv (|\Omega_{x}|^{2}+|\Omega_{y}|^{2})^2$, and $\Delta$ is the detuning. We further note that $\Omega_{y}$ is the Rabi frequency of the $y$-polarized pump field and $\Omega_{x}$ is the Rabi frequency of the $x$-polarized vacuum field. The homogeneous equation solved in cylindrical coordinates yields the LG family of solutions $u_{l,p}(\vec{r})$, given by Eq.~(\ref{eq:LGmodes}). We obtain the mode structure of the output beam by avoiding a numerical calculation and proceeding with a weak scattering approximation in the following way. We assume $\Omega_{x,y}$ on the right-hand side (r.h.s.) of Eq.~(\ref{eq:Rabi frequency propagation}) take the form of the input modes, i.e., $\Omega_{x,y} \to\varepsilon_{x,y}^0 u_{0,0}(\vec{r})$, $|\Omega|^{4} \to \varepsilon_{y}^{4}$, and use the fact that $\varepsilon_{x}^{0}\ll\varepsilon_{y}^{0}$ to simplify Eq.~(\ref{eq:Rabi frequency propagation}):
\begin{equation}\label{eq:1st order Rabi frequency propagation}
\begin{split}
 	 	\hat{L}\Omega_{x,y}= - u_{0,0}(\vec{r})\:\big[\kappa \gamma_{0}\:\dfrac{\varepsilon^{0}_{x,y}}{|\varepsilon_{y}^{0}|^{2}} \: u_{0,0}(\vec{r})^{*}u_{0,0}(\vec{r}) \\
 	+ i \frac{\kappa}{\Delta}\: \varepsilon_{x,y}^{0} \: (u_{0,0}(\vec{r})^{*}u_{0,0}(\vec{r}))^{2}\big].
\end{split}
\end{equation}
In this approximation we can regard the r.h.s. of Eq.~(\ref{eq:1st order Rabi frequency propagation}) as the source of the Rabi frequency on the left-hand side (l.h.s.) of Eq.~(\ref{eq:1st order Rabi frequency propagation}), and with this observation we define the appropriate sources $\rho_{x}$ and $\rho_{y}$ to simplify the notation, i.e., we transform Eq.~(\ref{eq:Rabi frequency propagation}) into $\hat{L}\Omega_{x,y} = \rho_{x,y}(\vec{r})$. We can now use a Green function method~\cite{Barton_book} to write an integral expression for $\Omega_{x,y}$:
  \begin{equation}
  \begin{split}
 	\Omega_{x,y}=\int r' dr' d\phi' \: K(\vec{r}\,|\vec{r}\,') \Omega_{x,y}^{\mathrm{homo}}(\vec{r}\,') \\
 	+ \int dz' \int r' dr' d\phi' \: G(\vec{r}\,|\vec{r}\,') \rho_{x,y}(\vec{r}\,'),
 	\end{split}
 \end{equation}
where $\Omega_{x,y}^{\mathrm{homo}}(\vec{r}\,') = \epsilon_{x,y}^{\mathrm{homo}}u_{0,0}(\vec{r})$ represents the portion of the beam passing through the cell unaltered, $K(\vec{r}|\vec{r}\,')$ is the propagator, and $G(\vec{r}|\vec{r}\,')$ is the Green function for Eq.~(\ref{eq:1st order Rabi frequency propagation}). An adjustable parameter, related to the absorption and properties of the cell, characterizes the relative strength of $\rho_{x,y}$ to $\Omega_{x,y}^{\mathrm{homo}}$. Thus, once $K(\vec{r}|\vec{r}\,')$ and $G(\vec{r}|\vec{r}\,')$ are known the problem is solved. The most effective way to write the propagator (and consequently the Green function) is in terms of the LG modes. We define the propagator and Green function accordingly:
\begin{equation}
\begin{split}
K(\vec{r}|\vec{r}\,') \equiv \sum_{l} \sum_{p} \: u_{l,p}^{*}(\vec{r}\,') \: u_{l,p}(\vec{r}),\\
G(\vec{r}|\vec{r}\,') \equiv  \Theta(z-z') K(\vec{r}\,|\vec{r}\,'),
\end{split}
\end{equation}
where $\Theta$ is the Heaviside step function.

We further utilize the LG modes with an expansion of the sources $\rho_{x,y}$:
  \begin{equation}
  \begin{split}
 	\rho_{x,y}(\vec{r}) = \sum_{l} \sum_{p} c_{l,p} \: u_{l,p}(\vec{r}),\\
 		c_{l,p}(z)=\int r dr d\phi \; u_{l,p}^{*}(\vec{r}) \: \rho_{x,y}(\vec{r}).
 	\end{split}
 \end{equation}
We pause to point out that, since the propagation equation inherits no $\phi$ dependence from either the atoms or the input fields, the beam solutions will be limited to $l=0$ modes. This is clearly seen through the integral for the $c_{l,p}$ coefficients by noting $\rho_{x,y}=\rho_{x,y}(r,z)$ and separating the $\phi$ phase from the LG mode:
\begin{equation}
  \begin{split}
 	c_{l,p}(z)=\int r dr  \rho_{x,y}(r,z)\:u_{l,p}^{*}(r,z)\int_{0}^{2\pi} d\phi\;e^{i l \phi}.
 	\end{split}
 \end{equation}
As one can see, the $\phi$ integral vanishes for $l\neq0$. The beauty of this approach is that, as a consequence of the orthogonality of the LG modes, we are able to find an analytic solution which retains the mode structure of the field. Using (8$\--$10) in (7) we arrive at the final solutions for the Rabi frequencies in the weak scattering limit:
  \begin{equation}
	  \label{eq:output_mode_solution}
  \begin{split}
 	\Omega_{x,y}= \Omega_{x,y}^{\mathrm{homo}}(\vec{r}) + \sum_{p} u_{0,p}(\vec{r}) \int_{\mathrm{cell}} dz' c_{0,p}(z').
 	\end{split}
 \end{equation}
One can perform these integrals with ease and our calculations show that the summation converges rapidly for $p\leq5$. Now, we proceed by introducing a new experimental investigation which can expose the complicated LG \textit{phase} structure predicted by the preceding theory, and hidden in the beam mode structure. 
\begin{figure}[h]
 \includegraphics[width=1.0\columnwidth]{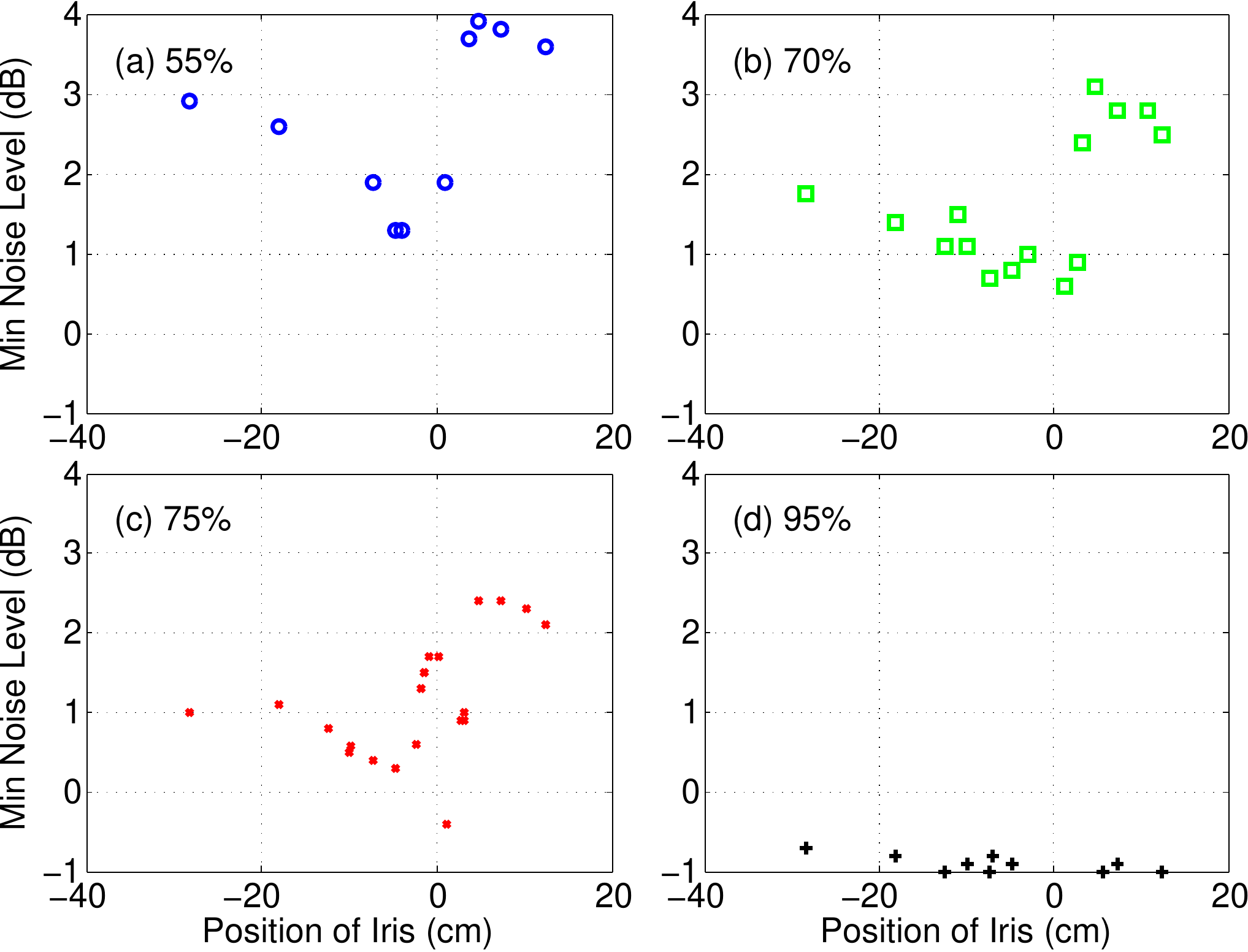}
  \caption{
          \label{fig:noise_vs_position}
          (Color online)
          Minimum quadrature noise  vs. a variable iris position.
	  The iris size was adjusted to maintain constant transmission
	  of the LO beam for each trace as depicted in the legend:
	  55\% (a, circles), 70\% (b, squares), 75\% (c, crosses), and 95\%
	  (d, plus signs).
          }
\end{figure}
\section{Spatial Properties of the Focused Output Beam}
\begin{figure}[b]
 \includegraphics[width=1.0\columnwidth]{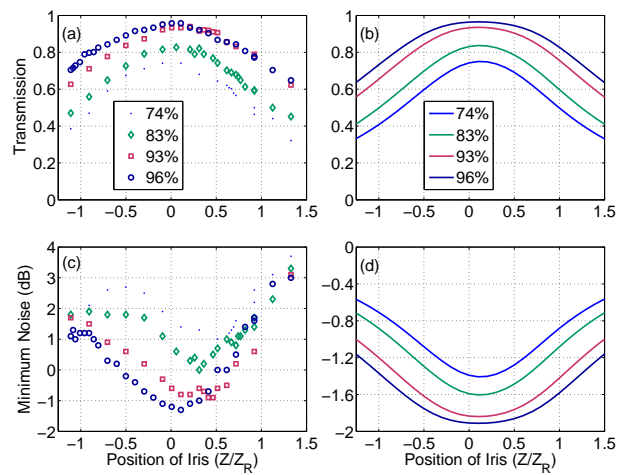}
  \caption{
          \label{fig:near_focus}
          (Color online)
	  Comparison of the experimental and theoretical (left and right
	  column) dependencies of the LO beam transmission (top row) and the
	  squeezed field minimal noise power (bottom row) on the iris
	  position for several fixed iris sizes.  The legend denotes the
	  peak transmission for each iris.
	  The Rayleigh range is $Z_{R}=$2.5~cm.
	  To calculate transmission and noise power, we used
	  Eq.~(\ref{eq:output_mode_solution}) and Eq.~(\ref{eq:beamsplitter}).
          }
\end{figure}
The phase factor $(2p+|l|+1)\arctan(z/z_{R})$ of the general LG mode (Eq.~(\ref{eq:LGmodes})) is referred to as the Gouy phase, and is responsible for a phase shift immediately around a beam focus. If our LO and squeezed field do in fact have a mixed LG structure, then we should observe asymmetries about such a focal point; otherwise, the beam will have a simple Gaussian profile on both sides of the focus. Referring back to Eq.~(\ref{eq:output_mode_solution}), we see that the semi-classical theory predicts a mixture of 5 different Gouy phases. Therefore, we built a one-to-one telescope to create an extra focal point and displaced the iris mask down the telescope to investigate the LG structure. The modified setup is depicted in Fig.~\ref{fig:setup}(b). As we moved the iris along the beam, we change its size to maintain the same transmission of the LO beam, i.e., we kept the same ratio of the iris radius and the $w(z)$. Therefore, our data is mostly effected by the different Gouy phase change in different spatial modes. We tracked the quantum noise in the maximally squeezed quadrature versus position of the iris. As one can see in Fig.~\ref{fig:noise_vs_position}, the quantum noise changes very drastically as the iris is moved right around the focal point. In the case that the iris size is large enough to allow high transmission, we see that the noise level does not change much as the iris is moved along the beam (see Fig.~\ref{fig:noise_vs_position}(d)). On the other hand, as transmission of the mask decreased, the noise in the squeezed quadrature went above shot noise (see Fig.~\ref{fig:noise_vs_position}(a, b, and c)).  Perhaps the most surprising is the noise dependence for the 75\% transmission mask (see Fig.~\ref{fig:noise_vs_position}(c)), where one can see a very sharp drop of the noise into the global minimum around $z$=1~cm, where it is squeezed below the shot-noise level. Similar but less pronounced behavior can be observed for the 70\% transmissive mask (see Fig.~\ref{fig:noise_vs_position}(b)).

For the same setup (see Fig.~\ref{fig:setup}(b)), we mapped the transmission and minimum noise curve vs. the position of irises with fixed diameters, as shown in Fig~\ref{fig:near_focus}. The transmission of LO exhibits some degree of asymmetry. The position of the maximum transmission shifted as we varied the iris size (see Fig.~\ref{fig:near_focus}(a)). For the irises with 93\% and 86\% peak transmission, there are more than one local minima. We use Eq.~(\ref{eq:output_mode_solution}) to predict the transmission level of the two LO field (see Fig.~\ref{fig:near_focus}(b)) and it is in a  qualitatively good agreement with our experimental data (see Fig.~\ref{fig:near_focus}(a)). We would like to again emphasize, that it is the interplay of Gouy phase shifts in the mode superposition (Eq.~(\ref{eq:output_mode_solution})) that create this peculiar dependence on the iris position around the focal point.

Based on the beam splitter model (Eq.~(\ref{eq:beamsplitter})), we calculate the expected minimum noise as in Fig.~\ref{fig:near_focus}(d). In the theoretical plot, the noise is always lower than shot noise since the model does not take into account the excess noise. But, we can still compare it with Fig.~\ref{fig:near_focus}(c) and find the overall behavior is quite similar. In both plots, the minimum in the noise power traces shift to the right, i.e., farther from the cell as the iris size shrinks. The qualitative agreement of these plots is further evidence of the LG structure of the squeezed vacuum beam. A more rigorous description necessarily requires a quantum treatment of both the light-matter interaction and the interaction with the spatial masks. Therefore, we next develop a simple second quantized theory that incorporates the LG structure of the semi-classical beam and predicts the distribution of the quantum noise for  the individual LG modes.    

\section{multi-mode Quantum Noise Calculations}
We now extend our previous semi-classical treatment of the multi-mode light-atom interactions to develop a second quantized theory that predicts the underlying modal structure of experimentally realized squeezing. At the first glance, one can think that our homodyne detection scheme is immune to any spatial mode structure modifications since the strong field, used as the local oscillator (LO), also passes through the vapor cell, so any spatial profile distortion are shared by both the squeezed optical field and the local oscillator. However, we still need to take into account the mode structure. Even if the overlap between squeezed modes and the local oscillator modes is perfect, distinct modes are going to contribute into the measured noise level differently. Therefore, to predict the resulting measured quantum noise, one must first predict the squeezing parameters for each of the vacuum mode, and then, using the semi-classical solution for the LO, simulate the results of the  homodyne detection. 

We proceed with the conventional single mode second quantization procedure, modifying it along the way to incorporate the higher-order LG mode structure. It is commonplace to represent the atomic polarization by a power expansion in the applied electric field strength $E$~\cite{Boyd_book}:
\begin{equation}\label{eq:NLpol}
	P= \sum_{n=1}^{\infty} P^{(n)} = \epsilon_{0}(\chi^{(1)}E + \chi^{(2)}E^{2} + \chi^{(3)}E^{3}+...).  
 \end{equation}
Typically, in the plane wave approximation, one would define the $n^{\mathrm{th}}$ order susceptibility as $\chi^{(n)}=P^{(n)}/\epsilon_{0}E^{n}$ ~\cite{Siegman_book}. However, in the paraxial approximation, special care must be taken since the electric field is confined to the beam axis. We reduce the electric field and polarization to $z$ dependence by integrating over a slice of the vapor cell perpendicular to the light beam. This is necessary, since the electric field dies off rapidly in $r$, and valid, since we are well within the validity of the paraxial approximation, i.e., each modes transverse profile simply rescales with $z$. Thus, from this point forward we are careful to distinguish the strictly $z$-dependent electric field and polarization 
\begin{equation}
\begin{split}
P(z)=\int_{\mathrm{cell}}rdrd\phi P(r,\phi,z)\\
E(z)=\int_{\mathrm{cell}}rdrd\phi E(r,\phi,z).
\end{split} 
 \end{equation}
Now that we have carefully chosen this representation, we may define the $n^{\mathrm{th}}$ order susceptibility as
\begin{equation}  
\chi^{(n)} \propto \dfrac{P^{(n)}(z)}{E^{n}(z)},
 \end{equation}
where $P^{(n)}(z)$ is the $n^{\mathrm{th}}$ order polarization in response of the input field $E(z)$. Then we further expand the optical polarization in terms of the LG modes: 
\begin{equation}
P^{(3)}(r,\phi,z)=\sum_{l,p}c_{l,p}u_{l,p},
 \end{equation}
where $c_{l,p}=\int rdrd \phi \: u^{*}_{l,p} \: P^{(3)}(r,\phi,z)$. Motivated by this expansion, we introduce the accompanying LG susceptibility for each LG mode:
\begin{equation} \label{eq:Lanningchi}
\chi^{(3)}_{l,p}(z) \equiv \dfrac{ P_{l,p}^{(3)}(z)}{E^{3}(z)},
 \end{equation}
 where we define the LG polarization $P_{l,p}^{(3)}(z)\equiv \int_{\mathrm{cell}}rdrd\phi\;c_{l,p}u_{l,p}$. In this representation, the LG susceptibility not only carries information about the LG spatial structure, it can be interpreted as the mechanism for ``cross-talk" between the Gaussian pump and the vacuum LG modes during the non-linear interaction. Therefore, we define an interaction Hamiltonian density following the prescription of Ref.~\cite{Drummond_book}:
\begin{equation}
\mathcal{H}_{l,p} \propto i \hbar \: \chi^{(3)}_{l,p}(z) \big(\hat{a}_{l,p}^{2} \hat{b}^{\dagger 2} - \hat{a}_{l,p}^{\dagger2} \hat{b}^{2} \big) 
 \end{equation}
where $\hat{a}_{l,p}$ is the $l,p$ spatial mode operator of the vacuum and $\hat{b}$  is the pump input mode. Thus, the full interaction Hamiltonian is 
\begin{equation}
\mathrm{H} \propto \int_{\mathrm{cell}} dz \: \sum_{l,p} \mathcal{H}_{l,p}.
 \end{equation}
We proceed by assuming the $y$-polarized pump mode is an undepleted coherent state and make the substitutions $\hat{b} \longrightarrow \beta e^{-i\omega_{p}t}$ and $\hat{b^{\dagger}} \longrightarrow \beta^{*} e^{i\omega_{p}t}$. This leads to the evolution operator
\begin{equation} \label{eq:Evolutionop}
\mathrm{U}= \exp \Big[ \int_{\mathrm{cell}} dz \sum_{l,p} (\eta^{*}_{l,p} t \; \hat{a}_{l,p}^{2} - \eta_{l,p} t \; \hat{a}_{l,p}^{\dagger 2}) \Big]
 \end{equation}
where $\eta_{l,p} \equiv \chi_{l,p}^{(3)} \beta^{2}$. Next, making the substitution
\begin{equation}
\xi_{l,p} \equiv 2 \int_{\mathrm{cell}} dz\; \eta_{l,p} t \equiv r_{l,p}e^{i \theta_{l,p}}  
 \end{equation}
we transform Eq.~(\ref{eq:Evolutionop}) into the form of the familiar ``squeezing'' operator with the addition of the LG indices~\cite{Gerry_and_Knight_book}:
\begin{equation} \label{eq:Squeezer}
\hat{S}(\xi) \equiv \mathrm{U}=
\exp \Big[ \sum_{l,p} \dfrac{1}{2} ( \xi^{*}_{l,p} \; \hat{a}_{l,p}^{2} - \xi_{l,p} \; \hat{a}_{l,p}^{\dagger 2}) \Big].
 \end{equation}
An inspection of Eq.(20) reveals that the behavior of the LG modes through the focus in the vapor cell drastically effects the distribution of squeezing among the modes. Likewise, the Gouy phase of each mode and the cell position/dimensions effectively determines the squeezing angle of each mode.    

After defining the proper squeezing operator, we can find the variances of the LG quadrature operators:
\begin{equation} 
 \begin{split}
  	\hat{X}_{1 \: l,p} = 
  	\dfrac{1}{2} (\hat{a}_{l,p} + \hat{a}^{\dagger}_{l,p}) \\
  		\hat{X}_{2 \: l,p} = 
  		\dfrac{1}{2i} (\hat{a}_{l,p} - \hat{a}^{\dagger}_{l,p})
	 \end{split}.
\end{equation}
The calculation is not particularly straightforward, but the result~\cite{Ma'90},  
\begin{equation}\label{eq:LGvariance}  
 \begin{split}
\big\langle(\Delta \hat{X}_{1,2 \: l,p})^{2} \big\rangle = 
\dfrac{1}{4} \Big[ \cosh^{2}r_{l,p} + \sinh^{2}r_{l,p} \\
\mp \: 2\sinh r_{l,p} \cosh r_{l,p}\cos\theta_{l,p} \Big]
	 \end{split},
\end{equation} 
is the familiar textbook result, but written for a particular LG mode~\cite{Gerry_and_Knight_book}. 
In other words, in this simplified scenario, each mode of squeezing has the statistics of a single squeezed mode. Furthermore, the photocurrent (difference) variance is a weighted sum of quadrature variances (some squeezed, some anti-squeezed) of all modes which overlap the LO~\cite{Bennink'02}. As we have shown previously, only the modes with $l=0$ are relevant to our experiment due to symmetry and consideration of angular momentum. Then one can show    
\begin{equation}\label{eq:QuadSum} 
 \begin{split}
\big\langle(\Delta i_{d})^{2} \big\rangle \propto \sum_{p} |O_{p}|^{2}[e^{-2r_{p}} \cos^{2}(\mathrm{Arg}~O_{p})\\
 + e^{2r_{p}} \sin^{2}(\mathrm{Arg}~O_{p}) ],
	 \end{split}
\end{equation} 
where we define the overlap integral $O_{p}$ as   
\begin{equation} 
 \begin{split}
O_{p} \equiv \dfrac{\int \Omega^{*}_{\mathrm{LO}} \phi_{p} \; d^{3}r }{\sqrt{\int |\Omega_{\mathrm{LO}}|^{2} \; d^{3}r}},
\end{split}
\end{equation} 
$\Omega_{\mathrm{LO}}$ is the local oscillator, and $\phi_{p}$ is the spatial function of the $p^{\mathrm{th}}$ mode (in this case simply the $u_{0,p}$ LG mode). To visualize these results we use the squeezed vacuum Wigner function; labeled here for each $l,p$ mode,  

\begin{equation}
	\begin{split}
		W_{l,p}(x,y)=\dfrac{2}{\pi}\exp\Big[ -(y^{2}+x^{2})\cosh r_{l,p}\\
+\big( (x^{2}-y^{2})\cos\theta_{l,p}+2xy\sin\theta_{l,p}\big)\sinh r_{l,p} \Big].
	\end{split}
\end{equation}

\newcommand{\subfigimg}[3][,]{%
  \setbox1=\hbox{\includegraphics[#1]{#3}}
  \leavevmode\rlap{\usebox1}
  \rlap{\hspace*{10pt}\raisebox{\dimexpr\ht1-2\baselineskip}{#2}}
  \phantom{\usebox1}
}
\begin{figure}[b]
  \centering
      \begin{tabular}{@{}p{0.5\linewidth}@{\quad}p{0.5\linewidth}@{}}
    \subfigimg[width=\linewidth]{(a)}{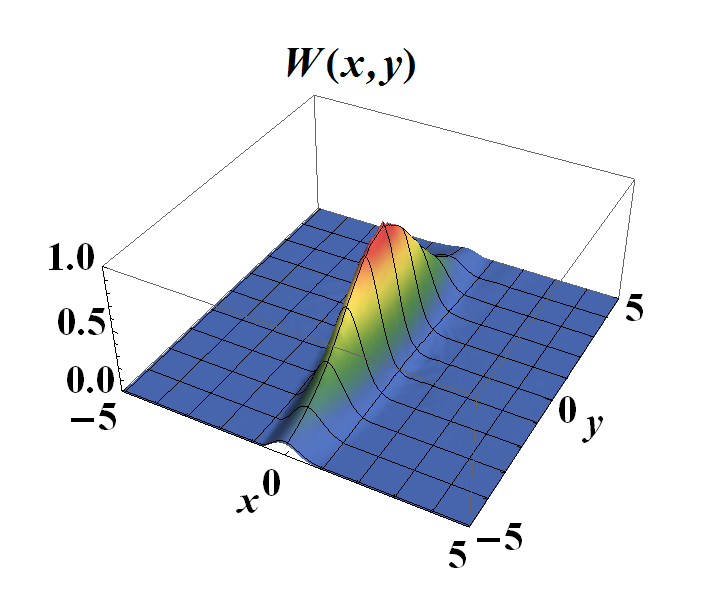} &
    \subfigimg[width=\linewidth]{}{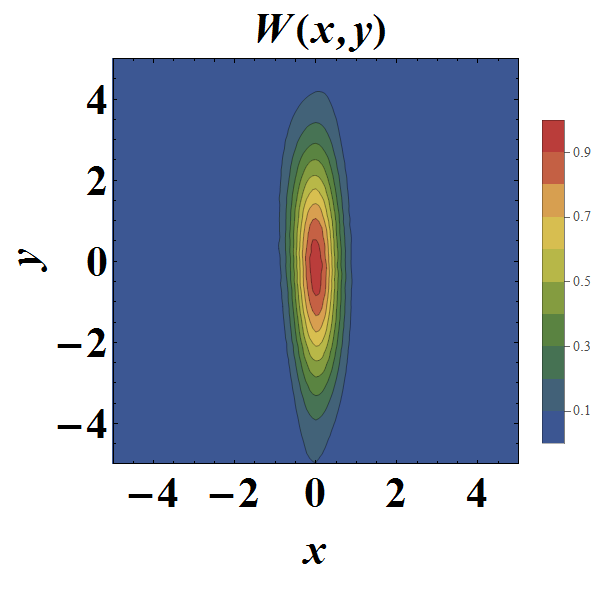} \\
    \subfigimg[width=\linewidth]{(b)}{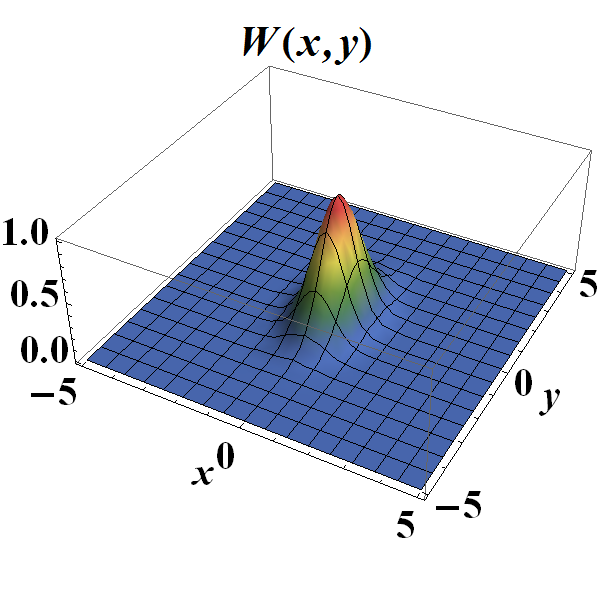} &
    \subfigimg[width=\linewidth]{}{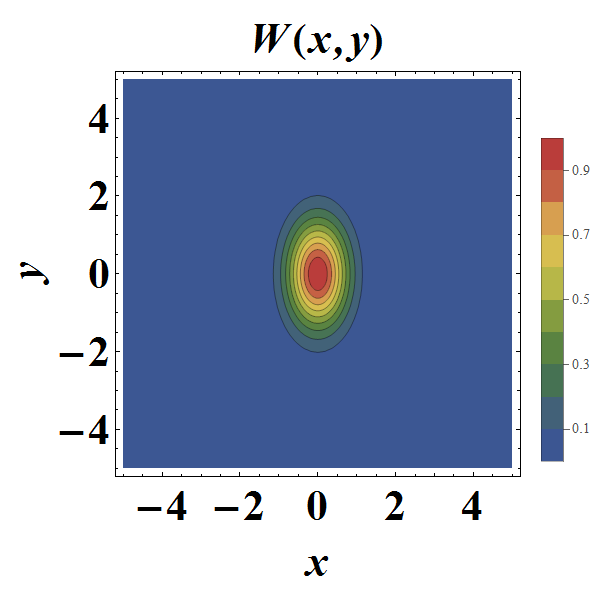} \\
  \end{tabular}
  \caption{Wigner function of (a) the experimentally realized squeezed state and (b) a hypothetical minimum uncertainty state with 1.9 dB of squeezing along the $X_{1}$ quadrature, shown here in 3D and contour. The simple second quantized theory predicts the multi-mode structure (Fig.~(\ref{fig:Qsqueezed})) that results in the same level of measured squeezing via homodyne detection. The axis labels x and y are proportional to the $X_{1}$ and $X_{2}$ quadratures respectively. The Wigner function has been rescaled by the peak amplitude of the vacuum Wigner function.} \label{fig:Qmodes}
\end{figure} 

Using quantum state tomography, we have reconstructed the Wigner function for our squeezed vacuum state (Fig.~(\ref{fig:Qmodes}a)). This Wigner function corresponds to noise suppression of $1.9\pm0.2$~dB below the SQL level  in the maximally squeezed quadrature  (the best suppression we were able to achieve in our experimental geometry), a modest amount of squeezing compare to the predictions of the single-mode theory~\cite{matsko_vacuum_2002,lezama_numerical_2008,mikhailov2011jmo}. However, the preceding theory suggests that more significant squeezing could be hiding in the spatial mode structure of the beam. To facilitate the comparison with the experimental results, we employ the second quantized theory to determine the value of the coupling constant $\kappa$ in Eq.\ref{eq:Rabi frequency propagation} so that the minimum quadrature noise of the theoretically calculated output measured via homodyne detection provides $1.9$~dB noise suppression in the $X_{1}$ quadrature, i.e., $\theta / 2 = 0$. The Wigner function of such a hypothetical state is plotted in Fig.~(\ref{fig:Qmodes}b). The same simple second quantized theory also gives us the decomposition of this state into various LG modes: first, we follow the preceding progression using the polarization derived in the semi-classical section (r.h.s. of Eq.~(\ref{eq:Rabi frequency propagation})) and the solution $\Omega_{y}$ (Eq.~(\ref{eq:output_mode_solution})) for the y-polarized beam since it is used as the LO during detection. Then, we re-scale the coupling constant such that the weighted sum of quadrature variances (Eq.~(\ref{eq:QuadSum})) reproduces the same squeezing of our hypothetical state. To take care of the scaling, we first recognize that since $\theta=0$, Eq.~(\ref{eq:LGvariance}) reduces to $\big\langle(\Delta \hat{X}_{1,2 \: l,p})^{2} \big\rangle = (1/4)e^{\mp2 r_{p}} $. Furthermore, in the limit of a strong LO, the photo-current difference variance is 
\begin{equation}\label{eq:QuadSum2} 
\big\langle(\Delta i_{d})^{2} \big\rangle \propto 4|\mathcal{E}_{\mathrm{LO}}|^{2}\big\langle(\Delta \hat{X}(\theta))^{2} \big\rangle
\end{equation}
where $\hat{X}(\theta)=(1/2)(\hat{a}e^{-i\theta}+\hat{a}^{\dagger}e^{i\theta})$ is the field quadrature operator at the angle $\theta$. Therefore, combining Eq.~(\ref{eq:QuadSum}) and Eq.~(\ref{eq:QuadSum2}) for the special case $\theta=0$, we impose that the scaling parameter $\gamma$ satisfy the equation
\begin{equation} 
 \begin{split}
|\mathcal{E}_{\mathrm{LO}}|^{2}e^{- 2 r_{\mathrm{exp}}} =\sum_{p} |O_{p}|^{2}[e^{-2 \, \gamma \, r_{p}} \cos^{2}(\mathrm{Arg~}O_{p})\\
 + e^{2 \, \gamma \, r_{p}} \sin^{2}(\mathrm{Arg}~O_{p}) ]
	 \end{split}
\end{equation}
where $r_{exp}\approx0.565$ is the experimental squeeze parameter and $|\mathcal{E}_{\mathrm{LO}}|^2 = \int rdrd\phi \, \Omega^{*}_{\mathrm{LO}}\Omega_{\mathrm{LO}}$.
We solve for $\gamma$ and determine the re-scaled LG squeeze parameters $r'_{p} = \gamma \; r_{p}$ and point out that the squeeze angles $\theta_{p}=\mathrm{Arg}(\xi_{p})$ remain the same. 

\begin{figure}[t]
  \centering
  \begin{tabular}{@{}p{0.5\linewidth}@{\quad}p{0.5\linewidth}@{}}
    \subfigimg[width=\linewidth]{(a)}{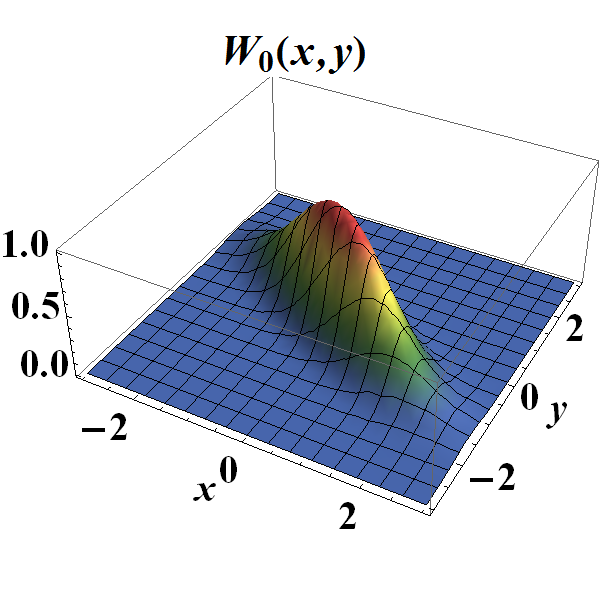} &
    \subfigimg[width=\linewidth]{}{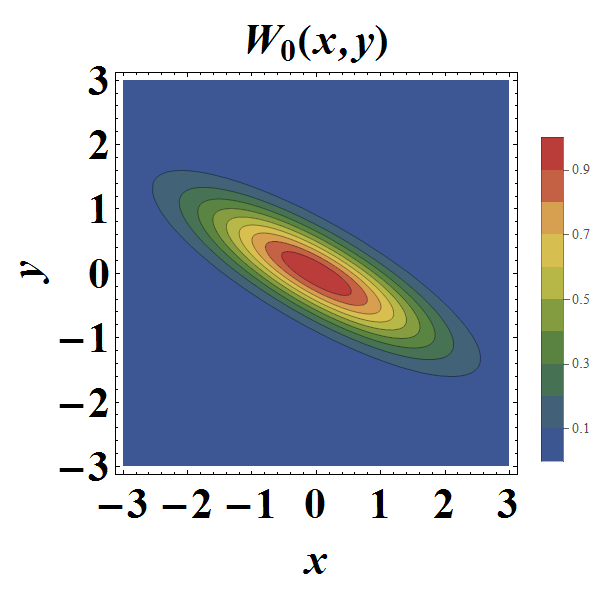} \\
    \subfigimg[width=\linewidth]{(b)}{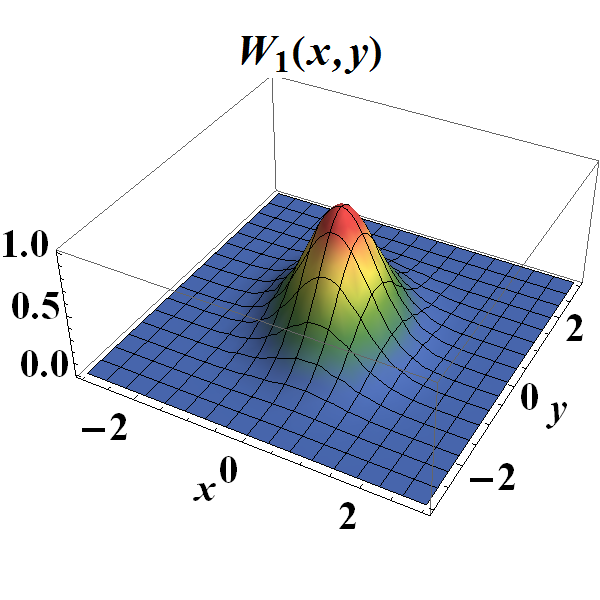} &
    \subfigimg[width=\linewidth]{}{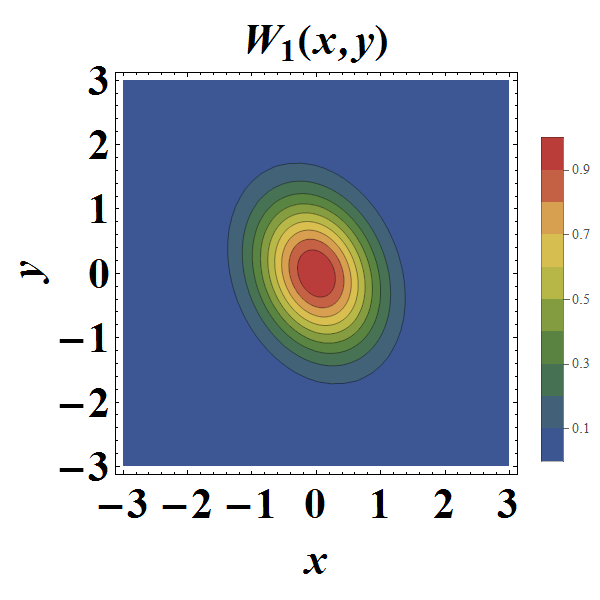} \\
    \subfigimg[width=\linewidth]{(c)}{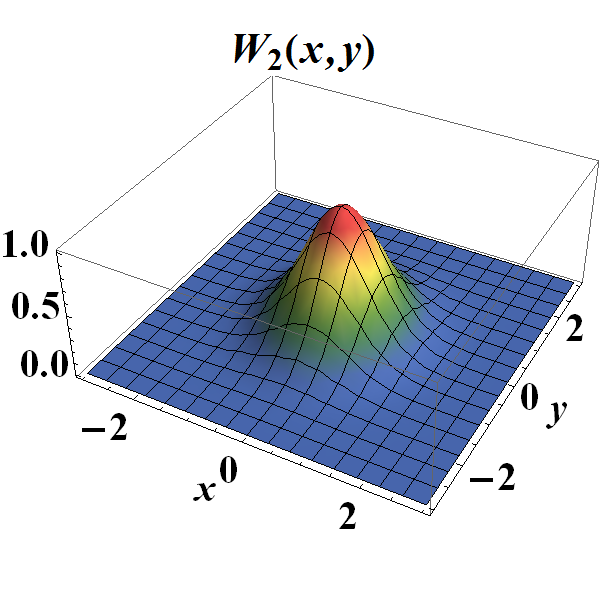} &
    \subfigimg[width=\linewidth]{}{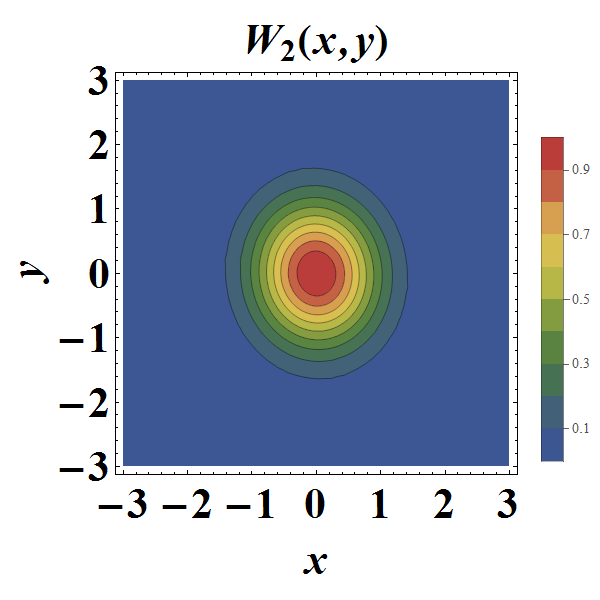} \\
  \end{tabular}
  \caption{The Wigner functions for the $p=0-2$ modes ((a),(b),(c) respectively) which, when measured simultaneously via homodyne detection, recreate the example Wigner function depicted in (Fig.~(\ref{fig:Qmodes}). The $p=3$ and greater modes are omitted since they appear essentially as vacuum modes. The axis labels x and y are proportional to the $X_{1}$ and $X_{2}$ quadratures respectively. The Wigner functions have been rescaled by the peak amplitude of the vacuum Wigner function.} \label{fig:Qsqueezed}
\end{figure}

\begin{table}[ht]
\caption{Squeezing Parameters for Various Modes} 
\centering 
\setlength{\tabcolsep}{10pt}
\begin{tabular}{ c | c c c c} 
\hline\hline 
$p$ & $r'_{p}$ & $\theta_{p}$/2 & $|O_{p}|$ & Arg($O_{p}$)\\ [0.5ex] 
\hline 
0 & 1.297 & $160^{\circ}$ & 0.995 & $71^{\circ}$ \\ 
1 & 0.315 & $113^{\circ}$ & 0.091 & $101^{\circ}$ \\
2 & 0.149 & $97^{\circ}$ & 0.031 & $123^{\circ}$ \\
3 & 0.029 & $25^{\circ}$ & 0.006 & $76^{\circ}$ \\
4 & 0.011 & $171^{\circ}$ & 0.004 & $38^{\circ}$ \\
5 & 0.010 & $18^{\circ}$ & 0.002 & $160^{\circ}$ \\ 
\hline 
\end{tabular}
\label{table:nonlin} 
\end{table}

Figure~(\ref{fig:Qsqueezed}) shows the Wigner functions for the squeezed LG modes, which comprise the hypothetical state, the $p=3$ and higher modes are omitted since they show no appreciable squeezing. Table~\ref{table:nonlin} summarizes the squeezing parameters and overlap integrals for different modes. It is important to point out that if only the fundamental mode $p=0$ could be isolated, its minimum quadrature noise would have been measured to be more that $11$~dB below the shot noise. The $p=1$ by itself (if isolated) would have displayed $-3$~dB of squeezing, but at the different quadrature angle, and the trend continues for higher modes, quickly approaching a coherent state.  Thus, we find that there is actually much more squeezing available in the individual modes, but in combination the measurable noise suppression is much worse due to the fact that the squeezing angles are out of phase. This deterioration is independent on the atomic excess noise ~\cite{lezama_numerical_2008,mikhailov2011jmo}, that is not taken into account in our simple theoretical model and that can further reduce the measured squeezing value. We hypothesize that the different Gouy phase for each mode rotates the squeezing angle for that mode at a different rate.

\section{Conclusion}
We demonstrated that the PSR squeezer generates a multi-spatial-mode squeezed field. We developed a semi-classical model, which qualitatively describes our experimental observation. Additionally, we present a simple second quantization procedure that shows how quadrature noise can be varied between LG spatial modes. Our results have application to improved single-mode squeezing and the production of squeezed light in higher-order modes upon demand.

The authors thank  G. Romanov, O. Wolfe, T. Horrom and N. Grewal for the assistance
with the experiment.  This research was supported by AFOSR grant FA9550-13-1-0098. In addition, R. N. L., Z. X. and J. P. D. acknowledge additional support from the ARO and the NSF, and the Northrop Grumman Corporation.

%



\begin{thebibliography}{30}%
\makeatletter
\providecommand \@ifxundefined [1]{%
 \@ifx{#1\undefined}
}%
\providecommand \@ifnum [1]{%
 \ifnum #1\expandafter \@firstoftwo
 \else \expandafter \@secondoftwo
 \fi
}%
\providecommand \@ifx [1]{%
 \ifx #1\expandafter \@firstoftwo
 \else \expandafter \@secondoftwo
 \fi
}%
\providecommand \natexlab [1]{#1}%
\providecommand \enquote  [1]{``#1''}%
\providecommand \bibnamefont  [1]{#1}%
\providecommand \bibfnamefont [1]{#1}%
\providecommand \citenamefont [1]{#1}%
\providecommand \href@noop [0]{\@secondoftwo}%
\providecommand \href [0]{\begingroup \@sanitize@url \@href}%
\providecommand \@href[1]{\@@startlink{#1}\@@href}%
\providecommand \@@href[1]{\endgroup#1\@@endlink}%
\providecommand \@sanitize@url [0]{\catcode `\\12\catcode `\$12\catcode
  `\&12\catcode `\#12\catcode `\^12\catcode `\_12\catcode `\%12\relax}%
\providecommand \@@startlink[1]{}%
\providecommand \@@endlink[0]{}%
\providecommand \url  [0]{\begingroup\@sanitize@url \@url }%
\providecommand \@url [1]{\endgroup\@href {#1}{\urlprefix }}%
\providecommand \urlprefix  [0]{URL }%
\providecommand \Eprint [0]{\href }%
\providecommand \doibase [0]{http://dx.doi.org/}%
\providecommand \selectlanguage [0]{\@gobble}%
\providecommand \bibinfo  [0]{\@secondoftwo}%
\providecommand \bibfield  [0]{\@secondoftwo}%
\providecommand \translation [1]{[#1]}%
\providecommand \BibitemOpen [0]{}%
\providecommand \bibitemStop [0]{}%
\providecommand \bibitemNoStop [0]{.\EOS\space}%
\providecommand \EOS [0]{\spacefactor3000\relax}%
\providecommand \BibitemShut  [1]{\csname bibitem#1\endcsname}%
\let\auto@bib@innerbib\@empty
\bibitem [{\citenamefont {Cerf}\ \emph {et~al.}(2007)\citenamefont {Cerf},
  \citenamefont {Leuchs},\ and\ \citenamefont {Polzik}}]{polzik_book}%
  \BibitemOpen
  \bibfield  {author} {\bibinfo {author} {\bibfnamefont {N.~J.}\ \bibnamefont
  {Cerf}}, \bibinfo {author} {\bibfnamefont {G.}~\bibnamefont {Leuchs}}, \ and\
  \bibinfo {author} {\bibfnamefont {E.~S.}\ \bibnamefont {Polzik}},\
  }\href@noop {} {\emph {\bibinfo {title} {Quantum Information with Continuous
  Variables of Atoms and Light}}}\ (\bibinfo  {publisher} {Imperial College
  Press, London},\ \bibinfo {year} {2007})\BibitemShut {NoStop}%
\bibitem [{\citenamefont {Boyer}\ \emph {et~al.}(2008)\citenamefont {Boyer},
  \citenamefont {Marino}, \citenamefont {Pooser},\ and\ \citenamefont
  {Lett}}]{lettSci08}%
  \BibitemOpen
  \bibfield  {author} {\bibinfo {author} {\bibfnamefont {V.}~\bibnamefont
  {Boyer}}, \bibinfo {author} {\bibfnamefont {A.~M.}\ \bibnamefont {Marino}},
  \bibinfo {author} {\bibfnamefont {R.~C.}\ \bibnamefont {Pooser}}, \ and\
  \bibinfo {author} {\bibfnamefont {P.~D.}\ \bibnamefont {Lett}},\ }\href
  {http://link.aps.org/doi/10.1103/PhysRevA.78.043816} {\bibfield  {journal}
  {\bibinfo  {journal} {Science}\ }\textbf {\bibinfo {volume} {321}},\ \bibinfo
  {pages} {544} (\bibinfo {year} {2008})}\BibitemShut {NoStop}%
\bibitem [{\citenamefont {Dowling}(2008)}]{Dowling2008}%
  \BibitemOpen
  \bibfield  {author} {\bibinfo {author} {\bibfnamefont {J.~P.}\ \bibnamefont
  {Dowling}},\ }\href {\doibase 10.1080/00107510802091298} {\bibfield
  {journal} {\bibinfo  {journal} {Contemp. Phys.}\ }\textbf {\bibinfo {volume}
  {49}},\ \bibinfo {pages} {125} (\bibinfo {year} {2008})}\BibitemShut
  {NoStop}%
\bibitem [{\citenamefont {Lassen}\ \emph {et~al.}(2007)\citenamefont {Lassen},
  \citenamefont {Delaubert}, \citenamefont {Janousek}, \citenamefont {Wagner},
  \citenamefont {Bachor}, \citenamefont {Lam}, \citenamefont {Treps},
  \citenamefont {Buchhave}, \citenamefont {Fabre},\ and\ \citenamefont
  {Harb}}]{PhysRevLett.98.083602}%
  \BibitemOpen
  \bibfield  {author} {\bibinfo {author} {\bibfnamefont {M.}~\bibnamefont
  {Lassen}}, \bibinfo {author} {\bibfnamefont {V.}~\bibnamefont {Delaubert}},
  \bibinfo {author} {\bibfnamefont {J.}~\bibnamefont {Janousek}}, \bibinfo
  {author} {\bibfnamefont {K.}~\bibnamefont {Wagner}}, \bibinfo {author}
  {\bibfnamefont {H.-A.}\ \bibnamefont {Bachor}}, \bibinfo {author}
  {\bibfnamefont {P.~K.}\ \bibnamefont {Lam}}, \bibinfo {author} {\bibfnamefont
  {N.}~\bibnamefont {Treps}}, \bibinfo {author} {\bibfnamefont
  {P.}~\bibnamefont {Buchhave}}, \bibinfo {author} {\bibfnamefont
  {C.}~\bibnamefont {Fabre}}, \ and\ \bibinfo {author} {\bibfnamefont {C.~C.}\
  \bibnamefont {Harb}},\ }\href {\doibase 10.1103/PhysRevLett.98.083602}
  {\bibfield  {journal} {\bibinfo  {journal} {Phys. Rev. Lett.}\ }\textbf
  {\bibinfo {volume} {98}},\ \bibinfo {pages} {083602} (\bibinfo {year}
  {2007})}\BibitemShut {NoStop}%
\bibitem [{\citenamefont {Molina-Terriza}\ \emph {et~al.}(2007)\citenamefont
  {Molina-Terriza}, \citenamefont {Torres},\ and\ \citenamefont
  {Torner}}]{Torres2007NaturePhy}%
  \BibitemOpen
  \bibfield  {author} {\bibinfo {author} {\bibfnamefont {G.}~\bibnamefont
  {Molina-Terriza}}, \bibinfo {author} {\bibfnamefont {J.~P.}\ \bibnamefont
  {Torres}}, \ and\ \bibinfo {author} {\bibfnamefont {L.}~\bibnamefont
  {Torner}},\ }\href {\doibase 10.1038/nphys607} {\bibfield  {journal}
  {\bibinfo  {journal} {Nature Physics}\ }\textbf {\bibinfo {volume} {3}},\
  \bibinfo {pages} {305} (\bibinfo {year} {2007})}\BibitemShut {NoStop}%
\bibitem [{\citenamefont {Fabre}(2008)}]{fabre:sfo-00270537}%
  \BibitemOpen
  \bibfield  {author} {\bibinfo {author} {\bibfnamefont {C.}~\bibnamefont
  {Fabre}},\ }\href {http://hal-sfo.ccsd.cnrs.fr/sfo-00270537} {\enquote
  {\bibinfo {title} {{Quantum Optics, from one mode to many modes}},}\ }
  (\bibinfo {year} {2008})\BibitemShut {NoStop}%
\bibitem [{\citenamefont {Berkhout}\ \emph {et~al.}(2010)\citenamefont
  {Berkhout}, \citenamefont {Lavery}, \citenamefont {Courtial}, \citenamefont
  {Beijersbergen},\ and\ \citenamefont {Padgett}}]{Berkhout2010PRLoamsorting}%
  \BibitemOpen
  \bibfield  {author} {\bibinfo {author} {\bibfnamefont {G.~C.~G.}\
  \bibnamefont {Berkhout}}, \bibinfo {author} {\bibfnamefont {M.~P.~J.}\
  \bibnamefont {Lavery}}, \bibinfo {author} {\bibfnamefont {J.}~\bibnamefont
  {Courtial}}, \bibinfo {author} {\bibfnamefont {M.~W.}\ \bibnamefont
  {Beijersbergen}}, \ and\ \bibinfo {author} {\bibfnamefont {M.~J.}\
  \bibnamefont {Padgett}},\ }\href {\doibase 10.1103/PhysRevLett.105.153601}
  {\bibfield  {journal} {\bibinfo  {journal} {Phys. Rev. Lett.}\ }\textbf
  {\bibinfo {volume} {105}},\ \bibinfo {pages} {153601} (\bibinfo {year}
  {2010})}\BibitemShut {NoStop}%
\bibitem [{\citenamefont {Pinel}\ \emph {et~al.}(2012)\citenamefont {Pinel},
  \citenamefont {Fade}, \citenamefont {Braun}, \citenamefont {Jian},
  \citenamefont {Treps},\ and\ \citenamefont {Fabre}}]{FabremultimodePRA2012}%
  \BibitemOpen
  \bibfield  {author} {\bibinfo {author} {\bibfnamefont {O.}~\bibnamefont
  {Pinel}}, \bibinfo {author} {\bibfnamefont {J.}~\bibnamefont {Fade}},
  \bibinfo {author} {\bibfnamefont {D.}~\bibnamefont {Braun}}, \bibinfo
  {author} {\bibfnamefont {P.}~\bibnamefont {Jian}}, \bibinfo {author}
  {\bibfnamefont {N.}~\bibnamefont {Treps}}, \ and\ \bibinfo {author}
  {\bibfnamefont {C.}~\bibnamefont {Fabre}},\ }\href {\doibase
  10.1103/PhysRevA.85.010101} {\bibfield  {journal} {\bibinfo  {journal} {Phys.
  Rev. A}\ }\textbf {\bibinfo {volume} {85}},\ \bibinfo {pages} {010101}
  (\bibinfo {year} {2012})}\BibitemShut {NoStop}%
\bibitem [{\citenamefont {Davis}\ \emph {et~al.}(1992)\citenamefont {Davis},
  \citenamefont {Gaeta},\ and\ \citenamefont {Boyd}}]{Davis:92}%
  \BibitemOpen
  \bibfield  {author} {\bibinfo {author} {\bibfnamefont {W.~V.}\ \bibnamefont
  {Davis}}, \bibinfo {author} {\bibfnamefont {A.~L.}\ \bibnamefont {Gaeta}}, \
  and\ \bibinfo {author} {\bibfnamefont {R.~W.}\ \bibnamefont {Boyd}},\ }\href
  {\doibase 10.1364/OL.17.001304} {\bibfield  {journal} {\bibinfo  {journal}
  {Opt. Lett.}\ }\textbf {\bibinfo {volume} {17}},\ \bibinfo {pages} {1304}
  (\bibinfo {year} {1992})}\BibitemShut {NoStop}%
\bibitem [{\citenamefont {Rochester}\ \emph {et~al.}(2001)\citenamefont
  {Rochester}, \citenamefont {Hsiung}, \citenamefont {Budker}, \citenamefont
  {Chiao}, \citenamefont {Kimball},\ and\ \citenamefont
  {Yashchuk}}]{budkerPRA01}%
  \BibitemOpen
  \bibfield  {author} {\bibinfo {author} {\bibfnamefont {S.~M.}\ \bibnamefont
  {Rochester}}, \bibinfo {author} {\bibfnamefont {D.~S.}\ \bibnamefont
  {Hsiung}}, \bibinfo {author} {\bibfnamefont {D.}~\bibnamefont {Budker}},
  \bibinfo {author} {\bibfnamefont {R.~Y.}\ \bibnamefont {Chiao}}, \bibinfo
  {author} {\bibfnamefont {D.~F.}\ \bibnamefont {Kimball}}, \ and\ \bibinfo
  {author} {\bibfnamefont {V.~V.}\ \bibnamefont {Yashchuk}},\ }\href {\doibase
  10.1103/PhysRevA.63.043814} {\bibfield  {journal} {\bibinfo  {journal} {Phys.
  Rev. A}\ }\textbf {\bibinfo {volume} {63}},\ \bibinfo {pages} {043814}
  (\bibinfo {year} {2001})}\BibitemShut {NoStop}%
\bibitem [{\citenamefont {Novikova}\ \emph {et~al.}(2002)\citenamefont
  {Novikova}, \citenamefont {Matsko},\ and\ \citenamefont
  {Welch}}]{novikova02JMO}%
  \BibitemOpen
  \bibfield  {author} {\bibinfo {author} {\bibfnamefont {I.}~\bibnamefont
  {Novikova}}, \bibinfo {author} {\bibfnamefont {A.~B.}\ \bibnamefont
  {Matsko}}, \ and\ \bibinfo {author} {\bibfnamefont {G.~R.}\ \bibnamefont
  {Welch}},\ }\href
  {http://search.ebscohost.com/login.aspx?direct=true&db=bth&AN=9019384&site=ehost-live}
  {\bibfield  {journal} {\bibinfo  {journal} {Journal of Modern Optics}\
  }\textbf {\bibinfo {volume} {49}},\ \bibinfo {pages} {2565 } (\bibinfo {year}
  {2002})}\BibitemShut {NoStop}%
\bibitem [{\citenamefont {Matsko}\ \emph {et~al.}(2002)\citenamefont {Matsko},
  \citenamefont {Novikova}, \citenamefont {Welch}, \citenamefont {Budker},
  \citenamefont {Kimball},\ and\ \citenamefont
  {Rochester}}]{matsko_vacuum_2002}%
  \BibitemOpen
  \bibfield  {author} {\bibinfo {author} {\bibfnamefont {A.~B.}\ \bibnamefont
  {Matsko}}, \bibinfo {author} {\bibfnamefont {I.}~\bibnamefont {Novikova}},
  \bibinfo {author} {\bibfnamefont {G.~R.}\ \bibnamefont {Welch}}, \bibinfo
  {author} {\bibfnamefont {D.}~\bibnamefont {Budker}}, \bibinfo {author}
  {\bibfnamefont {D.~F.}\ \bibnamefont {Kimball}}, \ and\ \bibinfo {author}
  {\bibfnamefont {S.~M.}\ \bibnamefont {Rochester}},\ }\href {\doibase
  10.1103/PhysRevA.66.043815} {\bibfield  {journal} {\bibinfo  {journal} {Phys.
  Rev. A}\ }\textbf {\bibinfo {volume} {66}},\ \bibinfo {pages} {043815}
  (\bibinfo {year} {2002})}\BibitemShut {NoStop}%
\bibitem [{\citenamefont {Ries}\ \emph {et~al.}(2003)\citenamefont {Ries},
  \citenamefont {Brezger},\ and\ \citenamefont
  {Lvovsky}}]{ries_experimental_2003}%
  \BibitemOpen
  \bibfield  {author} {\bibinfo {author} {\bibfnamefont {J.}~\bibnamefont
  {Ries}}, \bibinfo {author} {\bibfnamefont {B.}~\bibnamefont {Brezger}}, \
  and\ \bibinfo {author} {\bibfnamefont {A.~I.}\ \bibnamefont {Lvovsky}},\
  }\href {\doibase 10.1103/PhysRevA.68.025801} {\bibfield  {journal} {\bibinfo
  {journal} {Phys. Rev. A}\ }\textbf {\bibinfo {volume} {68}},\ \bibinfo
  {pages} {025801} (\bibinfo {year} {2003})}\BibitemShut {NoStop}%
\bibitem [{\citenamefont {Mikhailov}\ and\ \citenamefont
  {Novikova}(2008)}]{mikhailov2008ol}%
  \BibitemOpen
  \bibfield  {author} {\bibinfo {author} {\bibfnamefont {E.~E.}\ \bibnamefont
  {Mikhailov}}\ and\ \bibinfo {author} {\bibfnamefont {I.}~\bibnamefont
  {Novikova}},\ }\href {\doibase 10.1364/OL.33.001213} {\bibfield  {journal}
  {\bibinfo  {journal} {Opt. Lett.}\ }\textbf {\bibinfo {volume} {33}},\
  \bibinfo {pages} {1213} (\bibinfo {year} {2008})},\ \Eprint
  {http://arxiv.org/abs/0802.1558} {arXiv:0802.1558} \BibitemShut {NoStop}%
\bibitem [{\citenamefont {Barreiro}\ \emph {et~al.}(2011)\citenamefont
  {Barreiro}, \citenamefont {Valente}, \citenamefont {Failache},\ and\
  \citenamefont {Lezama}}]{lezama2011pra}%
  \BibitemOpen
  \bibfield  {author} {\bibinfo {author} {\bibfnamefont {S.}~\bibnamefont
  {Barreiro}}, \bibinfo {author} {\bibfnamefont {P.}~\bibnamefont {Valente}},
  \bibinfo {author} {\bibfnamefont {H.}~\bibnamefont {Failache}}, \ and\
  \bibinfo {author} {\bibfnamefont {A.}~\bibnamefont {Lezama}},\ }\href
  {\doibase 10.1103/PhysRevA.84.033851} {\bibfield  {journal} {\bibinfo
  {journal} {Phys. Rev. A}\ }\textbf {\bibinfo {volume} {84}},\ \bibinfo
  {pages} {033851} (\bibinfo {year} {2011})}\BibitemShut {NoStop}%
\bibitem [{\citenamefont {Agha}\ \emph {et~al.}(2010)\citenamefont {Agha},
  \citenamefont {Messin},\ and\ \citenamefont {Grangier}}]{grangier2010oe}%
  \BibitemOpen
  \bibfield  {author} {\bibinfo {author} {\bibfnamefont {I.~H.}\ \bibnamefont
  {Agha}}, \bibinfo {author} {\bibfnamefont {G.}~\bibnamefont {Messin}}, \ and\
  \bibinfo {author} {\bibfnamefont {P.}~\bibnamefont {Grangier}},\ }\href
  {\doibase 10.1364/OE.18.004198} {\bibfield  {journal} {\bibinfo  {journal}
  {Opt. Express}\ }\textbf {\bibinfo {volume} {18}},\ \bibinfo {pages} {4198}
  (\bibinfo {year} {2010})}\BibitemShut {NoStop}%
\bibitem [{\citenamefont {Horrom}\ \emph {et~al.}(2012)\citenamefont {Horrom},
  \citenamefont {Singh}, \citenamefont {Dowling},\ and\ \citenamefont
  {Mikhailov}}]{mikhailov2012sq_magnetometer}%
  \BibitemOpen
  \bibfield  {author} {\bibinfo {author} {\bibfnamefont {T.}~\bibnamefont
  {Horrom}}, \bibinfo {author} {\bibfnamefont {R.}~\bibnamefont {Singh}},
  \bibinfo {author} {\bibfnamefont {J.~P.}\ \bibnamefont {Dowling}}, \ and\
  \bibinfo {author} {\bibfnamefont {E.~E.}\ \bibnamefont {Mikhailov}},\ }\href
  {\doibase 10.1103/PhysRevA.86.023803} {\bibfield  {journal} {\bibinfo
  {journal} {Phys. Rev. A}\ }\textbf {\bibinfo {volume} {86}},\ \bibinfo
  {pages} {023803} (\bibinfo {year} {2012})},\ \Eprint
  {http://arxiv.org/abs/1202.3831} {arXiv:1202.3831} \BibitemShut {NoStop}%
\bibitem [{\citenamefont {Lezama}\ \emph {et~al.}(2008)\citenamefont {Lezama},
  \citenamefont {Valente}, \citenamefont {Failache}, \citenamefont
  {Martinelli},\ and\ \citenamefont {Nussenzveig}}]{lezama_numerical_2008}%
  \BibitemOpen
  \bibfield  {author} {\bibinfo {author} {\bibfnamefont {A.}~\bibnamefont
  {Lezama}}, \bibinfo {author} {\bibfnamefont {P.}~\bibnamefont {Valente}},
  \bibinfo {author} {\bibfnamefont {H.}~\bibnamefont {Failache}}, \bibinfo
  {author} {\bibfnamefont {M.}~\bibnamefont {Martinelli}}, \ and\ \bibinfo
  {author} {\bibfnamefont {P.}~\bibnamefont {Nussenzveig}},\ }\href
  {http://link.aps.org/abstract/PRA/v77/e013806} {\bibfield  {journal}
  {\bibinfo  {journal} {Phys. Rev. A}\ }\textbf {\bibinfo {volume} {77}},\
  \bibinfo {pages} {013806} (\bibinfo {year} {2008})}\BibitemShut {NoStop}%
\bibitem [{\citenamefont {Horrom}\ \emph {et~al.}(2011)\citenamefont {Horrom},
  \citenamefont {Lezama}, \citenamefont {Balik}, \citenamefont {Havey},\ and\
  \citenamefont {Mikhailov}}]{mikhailov2011jmo}%
  \BibitemOpen
  \bibfield  {author} {\bibinfo {author} {\bibfnamefont {T.}~\bibnamefont
  {Horrom}}, \bibinfo {author} {\bibfnamefont {A.}~\bibnamefont {Lezama}},
  \bibinfo {author} {\bibfnamefont {S.}~\bibnamefont {Balik}}, \bibinfo
  {author} {\bibfnamefont {M.~D.}\ \bibnamefont {Havey}}, \ and\ \bibinfo
  {author} {\bibfnamefont {E.~E.}\ \bibnamefont {Mikhailov}},\ }\href {\doibase
  10.1080/09500340.2011.594181} {\bibfield  {journal} {\bibinfo  {journal}
  {Journal of Modern Optics}\ }\textbf {\bibinfo {volume} {58}},\ \bibinfo
  {pages} {1936} (\bibinfo {year} {2011})},\ \Eprint
  {http://arxiv.org/abs/1103.1546} {arXiv:1103.1546} \BibitemShut {NoStop}%
\bibitem [{\citenamefont {Zhang}\ \emph {et~al.}(2013)\citenamefont {Zhang},
  \citenamefont {Soultanis}, \citenamefont {Novikova},\ and\ \citenamefont
  {Mikhailov}}]{mikhailov2013ol_vortex}%
  \BibitemOpen
  \bibfield  {author} {\bibinfo {author} {\bibfnamefont {M.}~\bibnamefont
  {Zhang}}, \bibinfo {author} {\bibfnamefont {J.}~\bibnamefont {Soultanis}},
  \bibinfo {author} {\bibfnamefont {I.}~\bibnamefont {Novikova}}, \ and\
  \bibinfo {author} {\bibfnamefont {E.~E.}\ \bibnamefont {Mikhailov}},\ }\href
  {\doibase 10.1364/OL.38.004833} {\bibfield  {journal} {\bibinfo  {journal}
  {Opt. Lett.}\ }\textbf {\bibinfo {volume} {38}},\ \bibinfo {pages} {4833}
  (\bibinfo {year} {2013})}\BibitemShut {NoStop}%
\bibitem [{\citenamefont {Marino}\ \emph {et~al.}(2008)\citenamefont {Marino},
  \citenamefont {Boyer}, \citenamefont {Pooser}, \citenamefont {Lett},
  \citenamefont {Lemons},\ and\ \citenamefont {Jones}}]{MarinoPhysRevLett08}%
  \BibitemOpen
  \bibfield  {author} {\bibinfo {author} {\bibfnamefont {A.~M.}\ \bibnamefont
  {Marino}}, \bibinfo {author} {\bibfnamefont {V.}~\bibnamefont {Boyer}},
  \bibinfo {author} {\bibfnamefont {R.~C.}\ \bibnamefont {Pooser}}, \bibinfo
  {author} {\bibfnamefont {P.~D.}\ \bibnamefont {Lett}}, \bibinfo {author}
  {\bibfnamefont {K.}~\bibnamefont {Lemons}}, \ and\ \bibinfo {author}
  {\bibfnamefont {K.~M.}\ \bibnamefont {Jones}},\ }\href {\doibase
  10.1103/PhysRevLett.101.093602} {\bibfield  {journal} {\bibinfo  {journal}
  {Phys. Rev. Lett.}\ }\textbf {\bibinfo {volume} {101}},\ \bibinfo {pages}
  {093602} (\bibinfo {year} {2008})}\BibitemShut {NoStop}%
\bibitem [{\citenamefont {Marino}\ \emph {et~al.}(2012)\citenamefont {Marino},
  \citenamefont {Clark}, \citenamefont {Glorieux},\ and\ \citenamefont
  {Lett}}]{MarinoEJPD2012}%
  \BibitemOpen
  \bibfield  {author} {\bibinfo {author} {\bibfnamefont {A.}~\bibnamefont
  {Marino}}, \bibinfo {author} {\bibfnamefont {J.}~\bibnamefont {Clark}},
  \bibinfo {author} {\bibfnamefont {Q.}~\bibnamefont {Glorieux}}, \ and\
  \bibinfo {author} {\bibfnamefont {P.}~\bibnamefont {Lett}},\ }\href {\doibase
  10.1140/epjd/e2012-30037-1} {\bibfield  {journal} {\bibinfo  {journal} {The
  European Physical Journal D}\ }\textbf {\bibinfo {volume} {66}},\ \bibinfo
  {eid} {288} (\bibinfo {year} {2012}),\
  10.1140/epjd/e2012-30037-1}\BibitemShut {NoStop}%
\bibitem [{\citenamefont {Siegman}(1986)}]{Siegman_book}%
  \BibitemOpen
  \bibfield  {author} {\bibinfo {author} {\bibfnamefont {A.}~\bibnamefont
  {Siegman}},\ }\href@noop {} {\emph {\bibinfo {title} {Lasers}}}\ (\bibinfo
  {publisher} {University Science Books, Sausalito CA},\ \bibinfo {year}
  {1986})\ pp.\ \bibinfo {pages} {103,647}\BibitemShut {NoStop}%
\bibitem [{\citenamefont {Corzo}\ \emph {et~al.}(2012)\citenamefont {Corzo},
  \citenamefont {Marino}, \citenamefont {Jones},\ and\ \citenamefont
  {Lett}}]{Lett2012PRLDoubleL}%
  \BibitemOpen
  \bibfield  {author} {\bibinfo {author} {\bibfnamefont {N.~V.}\ \bibnamefont
  {Corzo}}, \bibinfo {author} {\bibfnamefont {A.~M.}\ \bibnamefont {Marino}},
  \bibinfo {author} {\bibfnamefont {K.~M.}\ \bibnamefont {Jones}}, \ and\
  \bibinfo {author} {\bibfnamefont {P.~D.}\ \bibnamefont {Lett}},\ }\href
  {\doibase 10.1103/PhysRevLett.109.043602} {\bibfield  {journal} {\bibinfo
  {journal} {Phys. Rev. Lett.}\ }\textbf {\bibinfo {volume} {109}},\ \bibinfo
  {pages} {043602} (\bibinfo {year} {2012})}\BibitemShut {NoStop}%
\bibitem [{\citenamefont {Barton}(1989)}]{Barton_book}%
  \BibitemOpen
  \bibfield  {author} {\bibinfo {author} {\bibfnamefont {G.}~\bibnamefont
  {Barton}},\ }\href@noop {} {\emph {\bibinfo {title} {Elements of Green's
  Functions and Propagation}}}\ (\bibinfo  {publisher} {Oxford University
  Press, New York NY},\ \bibinfo {year} {1989})\BibitemShut {NoStop}%
\bibitem [{\citenamefont {Boyd}(2003)}]{Boyd_book}%
  \BibitemOpen
  \bibfield  {author} {\bibinfo {author} {\bibfnamefont {R.~W.}\ \bibnamefont
  {Boyd}},\ }\href@noop {} {\emph {\bibinfo {title} {Nonlinear Optics}}}\
  (\bibinfo  {publisher} {Academic Press},\ \bibinfo {year} {2003})\ pp.\
  \bibinfo {pages} {2,208}\BibitemShut {NoStop}%
\bibitem [{\citenamefont {Drummond}\ and\ \citenamefont
  {Ficek}(2004)}]{Drummond_book}%
  \BibitemOpen
  \bibfield  {author} {\bibinfo {author} {\bibfnamefont {P.~D.}\ \bibnamefont
  {Drummond}}\ and\ \bibinfo {author} {\bibfnamefont {Z.}~\bibnamefont
  {Ficek}},\ }\href@noop {} {\emph {\bibinfo {title} {Quantum Squeezing}}}\
  (\bibinfo  {publisher} {Springer-Verlag Berlin Heidelberg},\ \bibinfo {year}
  {2004})\ p.~\bibinfo {pages} {56}\BibitemShut {NoStop}%
\bibitem [{\citenamefont {Gerry}\ and\ \citenamefont
  {Knight}(2005)}]{Gerry_and_Knight_book}%
  \BibitemOpen
  \bibfield  {author} {\bibinfo {author} {\bibfnamefont {C.}~\bibnamefont
  {Gerry}}\ and\ \bibinfo {author} {\bibfnamefont {P.}~\bibnamefont {Knight}},\
  }\href@noop {} {\emph {\bibinfo {title} {Introductory Quantum Optics}}}\
  (\bibinfo  {publisher} {Cambridge University Press},\ \bibinfo {year}
  {2005})\ pp.\ \bibinfo {pages} {152,153}\BibitemShut {NoStop}%
\bibitem [{\citenamefont {Ma}\ and\ \citenamefont {Rhodes}(1990)}]{Ma'90}%
  \BibitemOpen
  \bibfield  {author} {\bibinfo {author} {\bibfnamefont {X.}~\bibnamefont
  {Ma}}\ and\ \bibinfo {author} {\bibfnamefont {W.}~\bibnamefont {Rhodes}},\
  }\href {http://dx.doi.org/10.1103/PhysRevA.41.4625} {\bibfield  {journal}
  {\bibinfo  {journal} {Phys.\ Rev.\ A}\ }\textbf {\bibinfo {volume} {41}},\
  \bibinfo {pages} {4625} (\bibinfo {year} {1990})}\BibitemShut {NoStop}%
\bibitem [{\citenamefont {Bennink}\ and\ \citenamefont
  {Boyd}(2002)}]{Bennink'02}%
  \BibitemOpen
  \bibfield  {author} {\bibinfo {author} {\bibfnamefont {R.~S.}\ \bibnamefont
  {Bennink}}\ and\ \bibinfo {author} {\bibfnamefont {R.~W.}\ \bibnamefont
  {Boyd}},\ }\href {\doibase 10.1103/PhysRevA.66.053815} {\bibfield  {journal}
  {\bibinfo  {journal} {Phys. Rev. A}\ }\textbf {\bibinfo {volume} {66}},\
  \bibinfo {pages} {053815} (\bibinfo {year} {2002})}\BibitemShut {NoStop}%
\end{thebibliography}
\end{document}